\documentclass[prb,aps,twocolumn,showpacs,10pt]{revtex4-1}
\usepackage{amsmath,amssymb,bm,graphicx,dsfont,color}
\usepackage[latin9]{inputenc}
\usepackage{amsmath}
\usepackage{slashed}
\usepackage{dsfont}
\usepackage{amstext}
\usepackage{amssymb}
\usepackage{amsbsy}
\usepackage{amsthm}
\usepackage{epsfig}
\usepackage{graphicx}
\usepackage{bbm}
\usepackage{hyperref}
\usepackage{color}
\usepackage{array}

\newcommand{\e}{\varepsilon}
\newcommand{\up}{\uparrow}
\newcommand{\down}{\downarrow}
\renewcommand{\>}{\rangle}
\renewcommand{\(}{\left(}
\renewcommand{\)}{\right)}

\renewcommand{\v}[1]{\mathbf{#1}} 

\begin{document}
 
\title{Classification of interacting electronic topological insulators in three dimensions}
\author{Chong Wang, Andrew C. Potter and T. Senthil}
\affiliation{Department of Physics, Massachusetts Institute of Technology,
Cambridge, MA 02139, USA}
 \date{\today}
\begin{abstract}


 A fundamental open problem in condensed matter physics is how the dichotomy 
between conventional and topological band insulators is modified in the presence 
of strong electron interactions. We show that there are 6 new electronic topological 
insulators that have no non-interacting counterpart. Combined with the previously 
known band-insulators, these produce a total of 8 topologically distinct phases. 
Two of the new topological insulators have a simple physical description as Mott 
insulators in which the electron spins form  spin analogs of the familiar topological 
band-insulator.  The remaining are obtained as combinations of these two `topological 
paramagnets' and the topological band insulator.   We prove that these 8 phases form a 
complete list of all possible interacting topological insulators, and are classified 
by a $\mathbb{Z}_2^3$ group-structure. Experimental signatures are also discussed for these phases.

\end{abstract}
\newcommand{\be}{\begin{equation}}
\newcommand{\ee}{\end{equation}}
\newcommand{\bea}{\begin{eqnarray}}
\newcommand{\eea}{\end{eqnarray}}
\newcommand{\p}{\partial}
\newcommand{\lp}{\left(}
\newcommand{\rp}{\right)}
\maketitle

The last few years have seen tremendous progress\cite{TI1,TI2,TI3,tenfold1,tenfold2} in our understanding of electronic topological insulators modeled by band theory. Despite this there is currently very little understanding of the  interplay between strong electron interactions and the phenomenon of topological insulation.  
Can interaction dominated phases be in a topological insulating state? Are there new kinds of topological insulators that might exist in interacting electron systems that have no non-interacting counterpart? These questions acquire particular importance in light of the ongoing experimental search for topological phenomena in strongly correlated materials with strong spin-orbit coupling.


It is important to  first distinguish the topological insulator  from a different class of more exotic topological phases - ones with a bulk gap but with ``intrinsic" topological order\cite{Wenbook} as exemplified most famously by the fractional quantum Hall states. Intrinsically topologically ordered phases have  exotic bulk excitations which may exhibit fractionalization of quantum numbers.


A fascinating {\em minimal} generalization of a topological insulator to interacting systems is to states of matter known as Symmetry Protected Topological (SPT) phases.  In contrast to more exotic generalizations\cite{note1,pesinlb} SPT phases have a bulk gap and no intrinsic topological order but nevertheless have  non-trivial surface states that are robust in the presence of a global internal symmetry.  

We focus here on the all-important example of time reversal symmetric insulating phases of electrons with a conserved global charge (corresponding to a global $U(1)$ symmetry).  Non-interacting insulators with this symmetry in $3D$ have a well known distinction\cite{FKM,TI1,TI2,TI3} between the topological and trivial band insulators (corresponding to a $\mathbb{Z}_2$ classification).   

We show that with interactions there are 6 other non-trivial topological insulating states corresponding to a classification by the group $\mathbb{Z}_2^3$. This group structure means that all these interacting topological insulators can be obtained from 3 `root' states and taking combinations. One of the 3 root states is the standard topological band insulator. The other two require interactions. They can be understood as Mott insulating states of the electrons where the resulting quantum spins have themselves formed an SPT phase. Such SPT phases of quantum spins were dubbed `topological paramagnets' in Ref. \onlinecite{avts12} and their properties in 3D elucidated. The three root states and their properties are  briefly described in Table. \ref{root}. 

A classic 1d example of a topological paramagnet is the Haldane spin-$1$ chain which has non-trivial end states that are protected by symmetry.  Substantial progress toward classification in diverse dimensions\cite{fidkit,1dsptclass1,1dsptclass2,1dsptclass3,chencoho2011sci,chencoho2011prb} has been reported. The physical properties of various such bosonic SPT phases in both 
two\cite{levingu,luav2012,tsml,liuxgw}  and in three dimensions\cite{avts12,xuts13,hmodl,burnellbc,metlitski,avdomain} have also been described in some detail, and provide crucial 
insights for the present work. 




 \begin{table*}[tttt]
\begin{tabular}{|>{\centering\arraybackslash}m{0.9in}|>{\centering\arraybackslash}m{1.5in}|>{\centering\arraybackslash}m{1in}|>{\centering\arraybackslash}m{1in}|}
\hline
{\bf Topological Insulator} &  {\bf Representative surface state} & {\bf $\mathcal{T}$-breaking transport signature} & {\bf $\mathcal{T}$-invariant  gapless superconductor} \\ \hline
Free fermion TI & Single Dirac cone & $\sigma_{xy}= \frac{\kappa_{xy}}{\kappa_0} =\pm1/2$ & None \\ \hline
Topological paramagnet I ($eTmT$) &  $\mathbb{Z}_2$ spin liquid with Kramers doublet spinon($e$) and vison($m$) & $\sigma_{xy} = \kappa_{xy} = 0$ & $N=8$ Majorana cones \\ \hline
Topological paramagnet II ($e_fm_f$) &  $\mathbb{Z}_2$ spin liquid with Fermionic spinon($e$) and vison($m$) & $\sigma_{xy} = 0; \frac{\kappa_{xy}}{\kappa_0}=\pm4$ & $N=8$ Majorana cones \\ \hline

\hline
\end{tabular}
\caption{Brief descriptions of the three fundamental non-trivial topological insulators, with their representative symmetry-preserving surface states, and surface signatures when either time-reversal or charge conservation is broken on the surface (with topological orders confined). $\sigma_{xy}$ is the surface electrical Hall conductivity in units of $\frac{e^2}{h}$. $\kappa_{xy}$ is the surface thermal Hall conductivity  and $\kappa_0 =  \frac{\pi^2}{3}\frac{k_B^2}{h} T$ ($T$ is the temperature). $N$ is the number of gapless Majorana cones protected by time-reversal symmetry when the surface becomes a superconductor. 
A combination of these measurements could uniquely determine the TI. 
}
\label{root}
\end{table*}%

Previous progress in understanding interacting electronic SPT phases is restricted to one\cite{fidkit,evelyn} and two\cite{luav2012,2D1,2D2,2D3,2D4} space dimensions. A formal abstract classification for some symmetries (which includes neither charge conservation, nor spin-$1/2$ electrons) in 3d has been attempted\cite{scoho} but leaves many physical questions unanswered.  
Our strategy - which sidesteps the difficulties of this prior approach - is to first constrain the symmetries and statistics of monopole sources of external electromagnetic fields. We then incorporate these constraints into a theory of the surface, and determine the resulting allowed distinct states.


In general it is natural to attempt to construct possible SPT phases of fermion system by first forming bosons as composites out of the fermions and putting the bosons in a bosonic SPT state.  However not all these boson SPTs remain distinct states in an electronic system. We determine that the distinct such states (see Appendix. \ref{S2}) can all be viewed as topological paramagnets as described above. 

While such spin-SPT phases can clearly exist, we give very general arguments that the only other electronic root state is the original topological band insulator.


We also clarify a number of other questions about interacting topological insulators (see end of the paper and Appendix. \ref{S6}, \ref{S7}).  For instance we explain the fundamental connection between topological insulation  and Kramers structure of the electron. 
 
\vspace{4pt}
\noindent{\bf Generalities}:  For any fully gapped insulator in 3D, the effective Lagrangian for an external electromagnetic field obtained by integrating out all the matter fields will take the form 
\begin{equation}
{\cal L}_{eff} = {\cal L}_{Max} + {\cal L}_\theta
\end{equation}
The first term is the usual Maxwell term and the second is the `theta' term:
\begin{equation}
{\cal L}_\theta = \frac{\theta}{4\pi^2} \v{E}\cdot\v{B}
\end{equation}
where $\v{E}$ and $\v{B}$ are the external electric and magnetic fields respectively. 

Under time reversal, $\theta \rightarrow - \theta$ and in a fermionic system the physics is periodic under $\theta \rightarrow \theta + 2\pi$. Time reversal symmetric insulators thus have $\theta = n \pi$ with $n$ an integer. Trivial time-reversal symmetric insulators have $\theta = 0$ while free fermion topological insulators have $\theta = \pi$\cite{qi}.  Any new interacting  TI that also has $\theta = \pi$ can be combined with the usual one to produce a TI with $\theta = 0$. Thus  it suffices to restrict attention to the possibility of new TIs which have $\theta = 0$. 

Consider the symmetry properties of monopole sources of the external magnetic field. At a non-zero $\theta$, this elementary monopole carries electric charge $\frac{\theta}{2\pi}$ so that it is neutral when  $\theta = 0$. Under time reversal the monopole becomes an anti-monopole as the magnetic field is odd. Formally if we gauge the global $U(1)$ symmetry to introduce a dynamical monopole field $m$ it must transform under time reversal as 
\begin{eqnarray}
\label{monopoleT}
{\cal T}^{-1} m {\cal T} & = & e^{i\alpha} m^\dagger \\
{\cal T}^{-1} m^\dagger {\cal T} & = & e^{-i\alpha} m 
\end{eqnarray}
However\cite{hmodl} (see Appendix. \ref{S1}) by combining with a gauge transformation we can set the phase $\alpha = 0$. 
Physically this is because the time reversed partner of a monopole lives in a different topological sector with opposite magnetic charge and hence is not simply a Kramers partner. 

This fixes the symmetry properties of the bulk monopole. There are still in principle two distinct choices corresponding to the statistics of the monopole: it may be either bosonic or fermionic. We will consider them in turn below. Bosonic monopoles will be shown to allow for the topological paramagnets mentioned above and nothing else.  Fermionic monopoles will be shown to not occur in electronic SPT phases.



\vspace{4pt}\noindent{\bf Topological insulators at ${\theta=0}$ - bosonic monopoles:} Consider the surface of any insulator with $\theta = 0$ and a bosonic monopole. 
This is conveniently incorporated into an effective theory of the surface  formulated in terms of degrees of freedom natural when the surface is superconducting, {\em i.e}, it spontaneously breaks the global $U(1)$ but not time reversal symmetry.  The suitable degrees of freedom then are $\frac{hc}{2e}$ vortices and (neutralized) Bogoliubov quasiparticles\cite{z2long}  (spinons) which have mutual semion interactions. In general we can also allow for co-existing topological order, i.e. other fractionalized quasi-particles, in the surface superconductor\cite{note2}.  This gives a dual description of $2D$ electronic systems that is particularly convenient to studying not just the superconducting phase but also some topologically ordered insulating phases. 



Imagine tunneling a monopole from the vacuum to the system bulk. Since the monopole is trivial in both regions, the tunneling event - which leaves a $\frac{hc}{e}$ vortex   on the surface - also carries no non-trivial quantum number.  Hence the surface dual effective field theory has a bosonic $\frac{hc}{e}$-vortex that carries no non-trivial quantum number. We can therefore proliferate (condense) the $\frac{hc}{e}$-vortex on the surface  which disorders the superconductor and  
 yields an insulator  with the full symmetry $U(1)\ltimes\mathcal{T}$ unbroken.  However as is well known from dual vortex descriptions\cite{bfn,z2long} of spin-charge separation in $2D$, the resulting state has intrinsic topological order.  



In  this surface topologically-ordered symmetry-preserving insulator, a quasi-particle of charge-$q$ sees the $\frac{hc}{e}$-vortex as a $2\pi q/e$ flux.  Hence, the $\frac{hc}{e}$-vortex condensate confines all particles with fractional charge and quantizes the charge to $q=ne$ for all the remaining particles in the theory (for a more detailed discussion  see Appendix. \ref{S3}). However, we can always remove integer charge from a particle without changing its topological sector by binding physical electrons. Hence the particle content of the surface topological order is $\{1,\epsilon,...\}\times\{1,c\}$, where only the physical electron $c$ in the theory is charged, and all the non-trivial fractional quasi-particles in $\{1,\epsilon,...\}$ are neutral. Since time-reversal operation preserves the $U(1)$ charge, its action has to be closed within the neutral sector $\{1,\epsilon,...\}$. We can therefore describe the surface topological order as a purely charge-neutral quantum spin liquid with topological order $\{1,\epsilon,...\}$, supplemented by the presence of a trivial electron band insulator, $\{1,c\}$. In particular, any gauge-invariant local operator made out of the topological theory must be neutral (up to binding electrons), but in an electron system a local neutral object has to be bosonic. Hence the theory should be viewed as emerging purely from a neutral boson system. This implies that the bulk SPT order should also be attributed to the neutral boson (spin) sector, {\em i.e}  it should be a SPT of spins in a Mott insulating phase of the electrons (a topological paramagnet).

The SPT states of neutral bosons with time-reversal symmetry are classified\cite{avts12,hmodl,burnellbc} by $\mathbb{Z}_2^2$, with two fundamental root non-trivial phases.  These can both be understood as Mott insulators in topological paramagnet phases. Adding to this the usual $\theta = \pi$ TI captured by band theory we have 3 root states corresponding to a $\mathbb{Z}_2^3$ classification. To establish that there are no other states we need to still consider the other possibility left open for the bulk response: a fermionic monopole.




\vspace{4pt}\noindent {\bf Topological insulators at ${\theta=0}$ - fermionic monopoles?}:
The possibility that the monopole may be fermionic in a system which also has fermionic charges is naively consistent with time-reversal symmetry.  However we can show that such a state cannot occur in any  electronic $3D$  SPT phase.    
Crucial to our argument is the requirement of `edgability' defined in Ref. \onlinecite{hmodl}. Any theory that can occur in strictly $d$-dimensions (as opposed to the surface of an SPT in $(d+1)$ dimensions) must admit a physical edge to the vacuum. We show that electronic systems with a fermionic monopole are not edgable.  

To illustrate the difficulty consider a Bose-Fermi mixture, with both the boson $b$ and the electron $c$ carrying charge-$1$. Now put the electron into a trivial band insulator, and the boson into a bosonic SPT state.  Then  the charge-neutral external monopole source becomes a fermion\cite{hmodl,metlitski}.   We may attempt to get rid of the bosons in the bulk by taking their charge gap to infinity ({\em i.e} projecting them out of the Hilbert space). However they will make their presence felt at the boundary and the theory is not edgable as a purely electronic system. Indeed we show in Appendix. \ref{S4} by a direct and general argument that fermionic statistics of the monopole in an SPT phase implies the existence of physical charge-$1$ bosons at the boundary.  This is not possible in a purely electronic system. 

\vspace{4pt}\noindent{\bf Physical characterization of interacting topological insulators}: We now describe phenomena which in principle can be used to completely experimentally identify the various TIs. We consider breaking symmetry at the surface to obtain states with no intrinsic topological order.   The results are summarized in Table.\ref{root}. 
A different, less practical, but conceptually powerful characterization is in terms of a gapped topologically ordered surface state which we describe in Appendix. \ref{S2}.

First consider surface states breaking time-reversal symmetry (and no intrinsic topological order). Of the 8 insulating phases we obtained, four have electromagnetic response $\theta = \pi$ (of which one is the topological band insulator) and four have $\theta = 0$ (of which one is the trivial insulator). The $\theta$ term in the response means that such a surface state will have quantized electrical Hall conductivity $\frac{e^2}{h}\nu$ with $\nu=\frac{\theta}{2\pi}+n$, where $n$ can be any integer signifying conventional integer quantum hall effect on the surface. A further distinction is obtained by considering the thermal Hall effect $\kappa_{xy}$ in this surface state. In general in a quantum Hall state $\kappa_{xy} = \nu_Q \frac{\pi^2}{3}\frac{k_B^2}{h} T$ where $k_B, T$ are Boltzmann's constant and the temperature respectively. The number $\nu_Q$ is a universal property of the quantum Hall state. 
 
Two of the $\theta = \pi$ insulators have $\nu_Q =\nu=1/2+n$ (including the topological band insulator) while the other two have $\nu_Q = \nu\pm 4$. Similarly two of the $\theta = 0$ insulators (including the trivial one) have $\nu_Q=\nu=n$ while the other two have $\nu_Q =\nu \pm 4$\cite{note3}. 
Thus a combined measurement of electrical and thermal Hall transport when ${\cal T}$ is broken at the surface can provide a very useful practical (albeit partial) characterization of these distinct topological insulators. 

Next we consider surface superconducting states (again without topological order) obtained by depositing an $s$-wave superconductor on top.   It was noticed in Ref. \onlinecite{TScSTO} that  the surface of the topological paramagnets I and II become identical to that of a  topological superconductor (see also Appendix. \ref{S5} for a simpler derivation). The corresponding free fermion superconductor has $N=8 ({\rm mod 16})$ gapless Majorana cones at the surface protected by time-reversal symmetry. Thus inducing  superconductivity on the surface of either Topological Paramagnet I or II leads to $8$ gapless Majorana cones which should be observable through photoemission measurements.  Taken together with the $T$-breaking surface transport we have a unique fingerprint for each of the 8 TIs.

\vspace{4pt}\noindent{\bf Other symmetries, Kramers fermions, and ${\theta = \pi}$ topological insulators}:
As a by-product of our considerations we are able to address a number of other fundamental questions about interacting topological insulators. 
For the free fermion systems the Kramers structure is what allows a topological insulator with $\theta = \pi$. What precise role, beyond free fermion band theory, does the Kramers structure of the electron play in enabling $\theta = \pi$ ?  We show non-perturbatively that any gapped  insulator with a $\theta = \pi$ response and no intrinsic topological order necessarily has charge carriers that are Kramers doublet fermions.  We also use a similar insight to show the necessity of magnetic ordering when the exotic bulk excitations of the topological Mott insulator phase of Ref. \onlinecite{pesinlb}
are confined. Finally we show that time reversal breaking electronic systems with global charge $U(1)$ symmetry have no interacting topological insulator phase in three dimensions. These results are described in Appendix. \ref{S6} and Appendix. \ref{S7}.

Our results set the stage for a number of future studies including identification of the new topological insulators in microscopic models and in real materials. 
Strongly correlated materials with strong spin orbit interactions are natural platforms for the various topological insulator phases we described.  We expect that our results will inform the many ongoing searches (e.g., in rare earth insulators, or in iridium oxides) for topological phenomena in such materials.

\textit{Acknowledgements - }  We thank X. Chen and C.-M. Jian, and particularly A. Vishwanath for useful discussions, and B. Swingle and P. A. Lee for comments on our manuscript.      This work was supported by Department of Energy DESC-8739- ER46872 (TS and CW), and and NSF Grant No. DGE-0801525 (ACP), and partially by the Simons Foundation by award number 229736 (TS).  TS also thanks the hospitality of Harvard University where this work was partially done. After this work was completed we learnt of Ref. \onlinecite{fSTO2} which also pointed out the relation between Kramers fermion and $\theta=\pi$ TI.

\appendix

\section{Time reversal action on the magnetic monopole}
\label{S1}
As the magnetic field is odd under time reversal, a magnetic monopole becomes an antimonopole. 
We briefly recapitulate the reasoning of Ref. \onlinecite{hmodl} to show that in Eq. (3) and (4) of the main paper the phase $\alpha$ can always be set to zero.  To see this we observe that the ${\cal T}$ operator can be combined with a (magnetic) gauge transformation to define a new time reversal operator: 
\begin{equation}
\label{gaugetr}
\tilde{\cal T} = U(\alpha) {\cal T}
\end{equation}
where $U(\alpha) = e^{-i\alpha q_m}$ where $q_m$ is the total magnetic charge. Since $q_m$ is odd under time-reversal, we have $U(\alpha)\mathcal{T}=\mathcal{T}U(\alpha)$, hence the order of product in Eq.~\eqref{gaugetr} does not matter. When acting on physical gauge invariant states $\tilde{\cal T}$ has the same effect as ${\cal T}$ but the monopole fields $m, m^\dagger$ transform with $\alpha = 0$. 

\section{Topologically ordered surface states}
\label{S2}
A powerful and complete characterization of the different three dimensional interacting topological insulators is in terms of a gapped symmetry preserving surface with intrinsic topological order.  The physical symmetries are realized in this surface topological order in a manner which cannot be realized in strict two dimensions. The surface topological order of the topological paramagnets was studied in Refs. \cite{avts12,hmodl,burnellbc}. The simplest such surface states have $\mathbb{Z}_2$ topological order, with two particles $e$ and $m$ having a mutual $\pi$-statistics. The Topological Paramagnet I  supports a surface theory in which both $e$ and $m$ particles are Kramers bosons (denoted as $eTmT$), while Topological Paramagnet II has a surface in which both $e$ and $m$ are non-Kramers fermions ($e_fm_f$). The third state, being a composite of the previous two, has $e$ and $m$ both being Kramers fermions ($e_fTm_fT$).

The topological band insulator can also be characterized in terms of its surface topological order. In contrast to the topological paramagnets the surface topological order in this case is non-abelian and such states have recently been studied in Ref. \onlinecite{fSTO1,fSTO2,fSTO3,fSTO4}. The resulting state are variants of the familiar Moore-Read state describing 
the $\nu=5/2$ quantum hall system, modified to accommodate time-reversal symmetry. In Table \ref{rootsto} we list the representative surface topological orders of the three root states described in the main text.

In hindsight, in interacting electron systems the descendants of neutral boson SPT states are naturally expected to arise. However, one could also have naively included the descendants of boson SPT states made out of Cooper pairs (charge-$2$ objects). The non-trivial boson SPT made out of physical bosons with charge $q=2$ supports a surface theory\cite{avts12,hmodl,metlitski} in which both $e$ and $m$ are non-Kramers bosons carrying charge $q/2=1$ (denoted as $eCmC$). However, since we have physical Kramers fermions with charge-$1$ in the system (the electrons), we can bind them with the $e$ and $m$ particles. This converts them to neutral Kramers fermions, which becomes exactly one of the SPT surface states ($e_fTm_fT$) of neutral bosons. Hence the SPT state made out of charge-$2$ bosons does not add any non-trivial fermion topological insulator. 

Apart from its conceptual value the study of the surface topological order  also provides a very useful theoretical tool to access the topological paramagnets. 
It allowed Ref. \onlinecite{hmodl} to construct the root states of the two time reversal symmetric topological paramagnets (as well as other bosonic SPT phases) in a system of coupled layers where each layer forms a state that is allowed in strictly $2d$ systems. Ref. \onlinecite{burnellbc} used the surface topological   order $e_fm_f$ (Topological Paramagnet II) to construct an exactly solvable model. While the constructions of Refs. \cite{avts12,hmodl,burnellbc} establish the existence of Topological Paramagnet II it is absent in the cohomology classification of Ref. \onlinecite{chencoho2011sci,chencoho2011prb}.  Understanding how to generalize the cohomology classification  to include this state is a challenge for the future.

 \begin{table}[tt]
\begin{tabular}{|>{\centering\arraybackslash}m{1.5in}|>{\centering\arraybackslash}m{1.8in}|}
\hline
{\bf Topological Insulator} &  {\bf Representative surface topological order} \\ \hline
Free fermion TI & Variants of Moore-Read state \\ \hline
Topological paramagnet I ($eTmT$) &  $\mathbb{Z}_2$ gauge theory with Kramers doublet $e$ and $m$, $\epsilon=em$ is singlet \\ \hline
Topological paramagnet II ($e_fm_f$) &  $\mathbb{Z}_2$ gauge theory with Fermionic $e$ and $m$, $\epsilon=em$ is also Fermionic  \\ \hline

\hline
\end{tabular}
\caption{Brief descriptions of the three fundamental non-trivial topological insulators, with their representative surface topological orders.}
\label{rootsto}
\end{table}%

 \section{Vortex condensate and charge quantization}
\label{S3}
Here we provide more details of the argument establishing that electronic topological insulators with $\theta = 0$ and a bosonic monopole can be reduced to bosonic topological paramagnets.  It is convenient to start with a symmetry preserving surface termination that has intrinsic topological order.  Such a surface state is characterized by a set of anyons $\{1,c, X, \bar{X}, Y_I\}$ where $I$ is a discrete label, and their corresponding braiding and fusion rules.  Each anyon will be characterized by a sharply quantized charge $q$ under the global $U(1)$ symmetry.  Let us denote this topological information and symmetry assignments as the initial surface anyon theory: $T_\text{initial}$.  

A useful theoretical device\cite{metlitski} is to consider creating a monopole source of an external (non-dynamical) magnetic field, and dragging that monopole through the topologically ordered surface at position $\v{R}$.  Such a monopole insertion event changes the external magnetic flux, $\Phi_B$, piercing the surface by $\frac{2\pi}{e}$ (in units where $\hbar=c=e=1$).  When the monopole sits close to the under-side of the surface, this extra flux, $\delta \Phi_B$, is concentrated in the vicinity of $\v{R}$.  Suppose we take a surface excitation, $Y$, with fractional charge $q_Y$, and drag it around a sufficiently large loop that encloses (nearly all) the additional magnetic-flux from the monopole insertion.  This process accumulates Berry phase $e^{2\pi i q_Y}\neq 1$ because of $Y$'s fractional charge.  However, the total monopole insertion event is a local physical process, and since there are no gapless excitations in the system it cannot have non-trivial action on distant events (clearly if $Y$ is arbitrarily far from the $\v{R}$, it should not be able to discern whether the monopole is infinitesimally above or infinitesimally below the surface).  Therefore, if $T_\text{initial}$ contains quasi-particles $Y_I$ with fractional charge, $q_{I}$, the monopole insertion event must also create a quasi-particle of type $X$ in the surface theory which has mutual statistics $\theta_{X,Y_I}=e^{-2\pi i q_{I}}$.  This mutual statistics then exactly compensates the non-trivial Berry phase from encircling the additional flux from the monopole insertion, and ensures that the overall monopole insertion event does not have unphysical non-local consequences.  Furthermore, since the bulk monopole is chargeless and bosonic, $X$, is a neutral boson.
 
We can similarly consider the time-reversed version of this process by inserting an anti-monopole from the vacuum into the bulk.  Let us denote by $\bar{X}$  the particle nucleated at the surface.  Clearly $X$ and $\bar{X}$ are exchanged by $\mathcal{T}$, indicating that, like $X$, $\bar{X}$ is a charge-neutral boson. The mutual statistics of an anyon $Y$ with $\bar{X}$ is then $e^{2\pi i q_Y}$. Further as the monopole and antimonopole can annihilate each other to give back the ground state $\bar{X}$ must be the antiparticle of $X$.  

These mutual statistics indicate that driving a phase transition in which $X,\bar{X}$ condense will confine all fractionally charged particles.  However, in general it is not guaranteed that the condensation of $X, \bar{X}$ preserves ${\cal T}$.  To avoid this issue, we take a detour through an intermediate superconducting phase in which descendants of $X,\bar{X}$ can be safely condensed while preserving $\mathcal{T}$.  This results in a topologically ordered state, $T_\text{final}$, which has the desired structure of a neutral boson theory.

Our strategy is to first enter a superconducting phase obtained by condensing the physical Cooper pair, $b\equiv c_\up c_\down$, from $T_\text{initial}$ and then to exit it through a different phase transition to reach the final topological order $T_\text{final}$. In the theory, $T_\text{initial}$, the Cooper pair is local with respect to all nontrivial anyons. Thus its condensation preserves the topological order $T_\text{initial}$.  The resulting topologically ordered superconductor is conventionally denoted SC$^*$ (see Ref. \onlinecite{z2long}) to distinguish it from the ordinary non-fractionalized BCS superconductor, SC.   

Let us denote the Cooper pair field by $b =\sqrt{\rho_b} e^{i\phi}$.  A long-wavelength effective Lagrangian density for the theory can be written:
\begin{align}\mathcal{L}[b,X,\bar{X},\cdots] &= \frac{\rho_b}{2}\(\partial_\mu \phi\)^2+\mathcal{L}_{T_\text{initial}}[X,\bar{X},Y_I,\dots]
\nonumber\\
&+\mathcal{L}_\text{mixed}[b,Y_I,\dots]
\end{align}
where $\mathcal{L}_{T_\text{initial}}[X,\bar{X},Y_I,\dots]$ is the Lagrangian density encoding the topological content of the topologically ordered phase, and $\mathcal{L}_\text{mixed} = \sum\lambda_{\{N_I\}} \prod_I\(e^{iq_{I}\phi/2}Y_I\)^{N_I}$ encodes all charge-conserving interaction terms between $b$ and gauge-invariant combinations of operators in the topologically ordered theory.  When $b$ condenses to obtain a superconducting phase, apart from the original topological quasiparticles, there will also be quantized vortex excitations where the phase $\phi$ of $b$ winds by $2n\pi$ with $n$ an integer.   Following the terminology of Ref. \onlinecite{z2long} we will call these vortons (to distinguish from the vortices of conventional superconductors without topological order). 

 We wish to disorder the superconducting order by condensing a suitable vortex  to obtain the desired insulating surface theory $T_\text{final}$. This may be done in a dual effective field theory in terms of the vorton degrees of freedom. To formulate such a dual field-theory, it is very convenient to introduce ``neutralized" fields: $\tilde{Y}_I = e^{iq_I \phi/2e} Y_I$, obtained by binding a fraction of the Cooper pair to $Y_I$.  In terms of these neutralized variables:
\begin{align} \mathcal{L} = \frac{\rho_b}{2}\(\partial_\mu \phi\)^2+\tilde{\mathcal{L}}[X,\bar{X},\tilde{Y}_I] \end{align}
The advantage of this choice of variables is now manifest, as the Cooper-pair phase $\phi$ is no longer directly coupled to the neutralized fields $\tilde{Y}_I$. 
The $\tilde{Y}_I$ however now acquire a phase $e^{\pi i q_I}$ on encircling an elementary vorton. 
 Following the standard duality transformation, we can re-write the boson current $j_b^\mu = \rho_b\partial_\mu\phi$ as the flux of a gauge-field $\alpha_\mu$: $j_b^\mu = \frac{\e^{\mu\nu\lambda}}{2\pi}\partial_\nu\alpha_\lambda$.  In the dual theory, the vorton field, denoted by $v$, is a bosonic field that couples minimally to this gauge field, and in addition has statistical interactions with the $\tilde{Y}$ particles:
\begin{align} 
\mathcal{L}_\text{dual} =& \frac{1}{8\pi^2\rho_b}\(\e^{\mu\nu\lambda}\partial_\nu\alpha_\lambda\)^2+ \frac{1}{2}|\(\partial_\mu-i\alpha_\mu-ia^{I}_\mu\)v|^2
\nonumber\\
&+V(|v|^2)+\tilde{\mathcal{L}}[X,\bar{X},\tilde{Y}_I]+\frac{\e^{\mu\nu\lambda}}{4\pi}a^I_\mu K_{IJ}\partial_\nu a^J_\lambda
\nonumber\\
&+\ell^{(I)}_Ja^J_\mu j_{Y_I}^\mu 
\end{align}
where the gauge fields, $a^I$, integer vectors $\ell^{(I)}$, and multi-component Chern-Simons term with K-matrix $K_{IJ}$ capture the mutual statistics between the vortons and the fields $Y_I$.  Here, $j_{Y_I}^\mu$ is the current of the $Y_I$ particles, and $V(|v|^2))$ is a potential for the vorton field.  

Now consider the particles $v^2X, \left(v^\dagger\right)^2 \bar{X}$. These carry vorticity $\pm 2$ and are interchanged under time reversal.  These are the relics of a monopole tunneling event in this superconducting state discussed in the main text.  Due to the coupling of $v$ to the dual gauge field, $\alpha_I$, we may always choose a gauge such that time reversal is implemented as: 
\begin{eqnarray}
{\cal T}^{-1} v^2 X {\cal T} &  = & \left(v^\dagger\right)^2 \bar{X}  \\
{\cal T}^{-1} \left(v^\dagger\right)^2 \bar{X} {\cal T} & = & v^2 X
\end{eqnarray}
 We may now condense $v^2X, \left(v^\dagger\right)^2 \bar{X}$ and preserve time reversal symmetry. The condensation also destroys the superconducting order and produces the desired new topological order $T_\text{final}$. Note that the neutralized particles $\tilde{Y}_I$ have no non-trivial mutual statistics with $v^2 X$ as the phase around the $v^2$ exactly cancels the phase around $X$. Hence they survive in $T_\text{final}$ as quasiparticles. The vortex condensate however quantizes electric charge to be an integer. In particular the charge $q$ bosons obtained by fractionalizing the Cooper pair $b_q = e^{\frac{iq \phi}{2}}$ are confined unless $q$ is an integer.   In effect the original electrically charged $Y_I$ particles are confined to the fractional bosons to produce the neutral $\tilde{Y}_I$ particles. 
 The vortons $v$ also survive as particles in $\text{final}$ but they are electrically neutral. 
 
 The detour through the superconductor essentially implements a `charge-anyon' separation of the original topological theory $T_\text{initial}$. This is completely analogous to the conceptual utility of superconducting degrees of freedom in implementing `spin-charge' separation in $2d$ insulators\cite{z2long}.  Though we will not elaborate this here an alternately route from $T_\text{initial}$ to 
 $T_\text{final}$ is through a parton construction where we fractionalize the charged anyons into a charged boson and  a neutral anyon.  
 
This proves that $T_\text{final}$ only has integer charged quasi-particles.  Without loss of generality, we may relabel the quasi-particle content of $T_\text{final}$ by binding an appropriate number of electrons to each quasi-particle to remove the remaining integer charge.  The resulting theory has quasi-particle content $\{1,v,\tilde{Y}_I\}\times\{1,c\}$, that can be decomposed into the direct product of a neutral boson sector $\{1,v,\tilde{Y}_I\}$ trivially accompanied by a gapped electron.  This completes the desired proof that the $\theta=0$ classification reduces to the classification of neutral bosonic phases.

\section{Impossibility of a Fermionic Monopole}
\label{S4}

In this section we provide a general argument against the possibility of fermionic monopoles in a purely electronic SPT insulator.  We will show that fermionic monopoles in the bulk necessarily leads to inconsistencies in the boundary theory, as long as the charge $U(1)$ symmetry is preserved.  When the charge $U(1)$ is gauged,  apart from monopoles we may also consider in the bulk dyons parametrized by $(q_m, q_e)$ where $q_e$ is the electric charge and $q_m$ the magnetic charge.   If the neutral monopole $(1,0)$ is fermionic in a purely electronic system (where the $(0,1)$ particle is identified with the electron) all dyons with $q_m = 1$ are also fermions.  If time reversal is broken in the bulk the $\theta$ value may change from $0$ leading to these dyons acquiring non-zero charge. However their statistics stays fermionic.  It follows that if any putative time reversal symmetric  electronic
topological insulator phase with a fermionic monopole exists then it  will stay a non-trivial topological insulator even in the absence of time reversal symmetry.   Thus it suffices to show that fermionic monopoles are forbidden in the absence of time reversal symmetry to rule out such putative topological insulators.

We will show that SPT states of electrons with a global $U(1)$ symmetry admit unphysical boundary excitations if the monopole is fermionic. 
Suppose we could construct a state with fermionic monopoles.  By the arguments of the previous section, we may describe this phase in terms of the surface topological order with particle content:
\begin{align} \{1,c,f,,Y_1,Y_2,\cdots\} \end{align}
Here, $f$ is the surface excitation corresponding to the bulk monopole, and hence is a neutral fermion having mutual statistics $e^{-2\pi i q_I/e}$, with particles $Y_I$ of charge $q_{I}$.
(Even if time reversal is not present we imagine tuning to a point where the monopole is neutral). 

Following an analogous line of reasoning in Appendix. \ref{S3}, we can now pair condense the remnant of the fermionic monopole $\<ff\>\neq 0$, which immediately confines all the fractionally charged particles $Y_I$ unless $q_{I}=ne/2$ for some integer $n$, due to their mutual statistics with $f$. By attaching enough physical electrons ($c$), we can always take the charge of the particles $Y_I$ to be either $0$ or $e/2$. The resulting theory can thus be written as:
\begin{align}
\label{tqftfm}
 \{1,c,f,C_I,N_I\} \end{align}
where $C_I$ have charge $e/2$, and $N_I$ are neutral quasi-particles. Note that $f$ is local with respect to $N_I$ and is a mutual semion with $C_I$. 



The neutral sector of the theory $\{1,f,N_I\}$ is closed under fusion and braiding due to charge conservation. Moreover they form a consistent topological field theory. To see this, let us momentarily dispense also with charge-conservation symmetry (for example by explicitly breaking it), and then condense $\<cf\>\neq 0$, which confines all 1/2-charged particles $C_I$ while keeping all the neutral particles $N_I$ unaffected.  Furthermore, as $f$ is local with respect to all the $N_I$'s, the theory $\{1,f,N_I\}$ can be viewed as a topological field theory of a system with physical fermion $f$ in the {\em absence} of any symmetry. Such a theory can then be confined to $\{1,f\}$ without obstruction. 

Returning to the original theory in Eqn \ref{tqftfm}  this implies that we may get rid of the neutral particles $N_I$ and be left with  \
\begin{align}\{1,c,f,C_i\} \label{eq:Inconsistent},\end{align} 
where $\{C_i\}$ is a subset of the original charge-$e/2$ particles $\{C_I\}$.


Without loss of generality, we can restrict our attention to a single species of fractional charge particle $C_1$, and its anti-particle.  The only possible fusion outcomes consistent with charge conservation are:  $C_1\times C_1\in \{c,cf\}$.  If two copies of $C_1$ fuse to $c$ then $c^\dagger C_1$ is the anti-particle of $C_1$.  However, this is not possible, since the topological spin (self-statistics) of $c^\dagger C_1$ and $C_1$ differs by $-1$, whereas anti-particles must have the same topological spin.  A similar argument rules out the possibility that two copies of $C_1$ fuses to $cf$.  

This line of reasoning shows that the topological order of Eq.~\ref{eq:Inconsistent} is internally inconsistent, unless there are no $C_i$ particles, i.e. unless the topological order contains only the following particles:
\begin{align} \{1,c,f\}. \end{align}
Since $f$ has trivial mutual statistics with $c$, it must be a physical object that is microscopically present in the system (i.e. is not an emergent particle).  However, there is no such neutral fermion degree of freedom in an electronic system.   It follows that in a purely electronic system the monopole cannot be fermionic in an SPT phase with global $U(1)$ 
symmetry. 

We note that the Bose-Fermi example constructed in the main paper has a neutral fermion excitation ( a bound state of the boson and fermion) and hence is allowed to have a fermionic monopole. Let us examine this more closely.  We put the electron into a trivial band insulator, and the boson into a boson  topological insulator.  Then  the charge-neutral external monopole source becomes a fermion\cite{hmodl,metlitski}.  We initially consider such a system in a geometry with no boundaries. We then tune the boson charge gap to infinity, so that the charged bosons disappear from the spectrum, and we are left with a purely electronic theory. But since the fermionic monopole does not carry any boson charge, it survives as the only charge-neutral monopole. Now the bulk theory is exactly what we were looking for, but we need to examine its boundary and see if it is consistent with a time-reversal invariant electronic system.  

As the electrons are in a trivial insulator they do not contribute anything special on the boundary, so we only have to worry about surface states of the $eCmC$ boson SPT. We first consider a symmetric surface state with topological order. It is known\cite{avts12} that one of the possible surface states of the bosonic TI is described by a $\mathbb{Z}_2$ gauge theory with both $e$ and $m$ carrying charge-$1/2$ and the $\epsilon$ fermion being charge-neutral (the state denoted $eCmC$ in Ref. \onlinecite{hmodl}). By setting the boson charge-gap to infinity, the $e$ and $m$ particles disappear from the spectrum, but the neutral $\epsilon$ fermion survives as a gauge-invariant local object, which is not allowed in a system purely made of charged fermions. Another way to see the inconsistency of the surface is to look at the surface state without topological order in which time-reversal symmetry is broken. The boson topological insulator leads to a surface electrical quantum hall conductance $\sigma_{xy}=\pm1$ and thermal hall conductance $\kappa_{xy}=0$.\cite{avts12} The difference of $\sigma_{xy},\kappa_{xy}$ between the two time-reversal broken states should correspond to an electronic state in two dimensions without topological order. Here we have $\Delta\sigma_{xy}=2$ and $\Delta\kappa_{xy}=0$, which cannot be realized from a purely electronic system without topological order. Indeed adding integer quantum Hall states of electrons increases $\sigma_{xy}, \kappa_{xy}$ by the same amount. It is possible to add a neutral boson integer quantum Hall state without topological order but that requires $\sigma_{xy = 0}, \kappa_{xy} = 0 (mod 8)$. Hence the boundary as a purely electronic theory is not consistent with time-reversal symmetry, and the bulk theory cannot be realized in strict three dimensions, although it may be realizable at the surface of a four dimensional system.  We also note that if we allow topological (or other exotic long range entanglement) in the bulk then the monopole may be fermionic.

\section{Equivalence between $N=8$ Majorana cones and the $eTmT$ topological order}
\label{S5}

In this section, we provide a physical construction of the $eTmT$ topological order from the $N=8$ Majorana-cone surface state of a time-reversal invariant topological superconductor phase.  We start from the free theory
\begin{equation}
\mathcal{L}_{free}=\sum_{i=1}^{8}\chi^{T}_{i,a}(p_x\sigma^x+p_y\sigma^z)_{a,b}\chi_{i,b}
\end{equation}
where $i\in \{1,\dots ,4\}$ and $a \in \{\up,\down\}$, and with time reversal acting on the real (Majorana) fermions as
\begin{equation}
\mathcal{T}\chi_i\mathcal{T}^{-1}=i\sigma^y_{ab}\chi_{i,b}.
\end{equation}

We can group the theory into four complex (Dirac) fermions by writing
\begin{equation}
\psi_{i,a}=\chi_{2i,a}+i\chi_{2i-1,a},
\end{equation}
the Lagrangian then simply describes four gapless Dirac cones
\begin{equation}
\mathcal{L}_{free}=\sum_{i=1}^{4}\psi^{\dagger}_i(p_x\sigma^x+p_y\sigma^z)\psi_i,
\end{equation}
in which time-reversal acts as
\begin{equation}
\mathcal{T}\psi_i\mathcal{T}^{-1}=i\sigma_y\psi^{\dagger}_i.
\end{equation}

It is easy to see that the theory is protected from gap-opening at the free (quadratic) level. We can then ask, could a non-perturbative gap be opened when interaction is introduced? The way to tackle this problem is to first introduce a symmetry-breaking mass term into the fermion theory, viewing the mass term as an fluctuating order parameter, and ask if one can recover the symmetry by disordering the phase of the mass field.

For this purpose it is convenient to first introduce an auxiliary global $U(1)$ symmetry
\begin{equation}
U_{\theta}\psi U^{-1}_{\theta}=e^{i\theta}\psi
\end{equation}
as a microscopic symmetry in the model (rather than a subgroup of the emergent $SO(8)$ flavor symmetry). This auxiliary symmetry will be removed at the end of the argument, so the final result does not depend on the existence of this $U(1)$ symmetry.

The total symmetry is now enlarged to $U(1)\times\mathcal{T}$, with $U_{\theta}\mathcal{T}=\mathcal{T}U_{\theta}$ (i.e. the conserved quantity associated with the auxiliary $U(1)$ symmetry changes sign under $\mathcal{T}$ like a component of spin rather than an electrical charge). One can now write down a pairing-gap term into the theory
\begin{equation}
\label{gap}
\mathcal{L}_{gap}=i\Delta\sum_{i=1}^{4}\psi_i\sigma_y\psi_i+h.c.
\end{equation}
which breaks both $U(1)$ and $\mathcal{T}$ ($\Delta\to-\Delta^*$ under time-reversal because $\mathcal{T}^2=-1$ on physical fermions). The task for us now is then to disorder the field $\Delta$ and restore time-reversal symmetry. The virtue of the auxiliary $U(1)$ symmetry shows up here: the field $\Delta$ is $XY$-like, so to disorder it we can follow the familiar and well-understood route of proliferating vortices of the order parameter

It is important here to notice that although the gap in Eq.~\eqref{gap} breaks both $U(1)$ and $\mathcal{T}$, it does preserve a time-reversal-like subgroup generated by $\tilde{\mathcal{T}}=\mathcal{T}U_{\pi/2}$. Since we want to restore $\mathcal{T}$ by disordering $\Delta$ (which surely will restore $U(1)$), we must do it while preserving $\tilde{\mathcal{T}}$. This ``modified time-reversal" looks almost like the original one, but there is a crucial difference: $\tilde{\mathcal{T}}^2=1$ when acting on the fermion field $\psi$.

Now we are ready to disorder the field $\Delta$. At first glance it seems sufficient just to proliferate the fundamental vortex ($hc/2e$-vortex) and obtain a trivial gapped insulator. However, as we will see below, $\tilde{\mathcal{T}}^2=-1$ on these fundamental vortices, hence proliferating them could not restore time-reversal symmetry. 

The vortex here is subtle because of the fermion zero-modes associated with it. It is well-known that a superconducting Dirac cone gives a Majorana zero-mode in the vortex core\cite{fkmajorana}. So the four Dirac cones in total gives two complex fermion zero-modes $f_{1,2}$. We then define different vortex operators as
\begin{equation}
v_{nm}|GS\rangle=\left(f_1^{\dagger}\right)^n\left(f_2^{\dagger}\right)^m|FN\rangle,
\end{equation}
where $|FN\rangle$ denotes the state with all the negative-energy levels filled in a vortex background. The $U(1)$ being spin-like under $\mathcal{T}$ (hence $\tilde{\mathcal{T}}$ also) means that a vortex configuration is time-reversal invariant. The only non-trivial action of $\tilde{\mathcal{T}}$ is thus on the zero-modes:
\begin{equation}
\tilde{\mathcal{T}}f_{1,2}\tilde{\mathcal{T}}^{-1}=f^{\dagger}_{1,2},
\end{equation}
and by choosing a proper phase definition:
\begin{equation}
\tilde{\mathcal{T}}|FN\rangle=f^{\dagger}_1f^{\dagger}_2|FN\rangle.
\end{equation}
It then follows straightforwardly that $\{v_{00},v_{11}\}$ and $\{v_{01},v_{10}\}$ form two "Kramers" pairs under $\tilde{\mathcal{T}}$. Moreover, since the two pairs carry opposite fermion parity, they actually see each other as mutual semions.

We thus conclude that to preserve the symmetry, the ``minimal" construction is to proliferate double vortices. The resulting insulating state has $\mathbb{Z}_2$ topological order $\{1,e,m,\epsilon\}$ with the $e$ being the remnant of $\{v_{00},v_{11}\}$, $m$ being the remnant of $\{v_{01},v_{10}\}$, and $\epsilon$ is the neutralized fermion $\tilde{\psi}$.

Now the full $U(1)\times\mathcal{T}$ is restored, we can ask how are they implemented on $\{1,e,m,\epsilon\}$. Obviously these particles are charge-neutral, so the question is then about the implementation of $\mathcal{T}$ alone. However, since the particles are neutral the extra auxiliary $U(1)$ rotation in $\tilde{\mathcal{T}}$ is irrelevant and they transform identically under $\tilde{\mathcal{T}}$ and $\mathcal{T}$. Hence we have $\mathcal{T}^2=\tilde{\mathcal{T}}^2=-1$ on $e$ and $m$, and $\mathcal{T}^2=\tilde{\mathcal{T}}^2=1$ on $\epsilon$, which is exactly the topological order $eTmT$. The charged physical fermion $\psi$ is now trivially gapped and plays no role in the topological theory, one can thus introduce explicit pairing to break the auxiliary $U(1)$ symmetry. Since topological order stems from the charge-neutral sector, pair-condensation of $\psi$ does not alter the topological order, and the resulting state is just the $eTmT$ state with only $\mathcal{T}$ symmetry.

\section{Spinless fermions and other symmetries}
\label{S6}

We first provide the proof that  a $\theta = \pi$ electromagnetic response in a time reversal invariant insulator implies that the charge  carriers are Kramers fermions. 

When the global $U(1)$ symmetry is gauged, $\theta = \pi$ implies that the monopoles of the resulting $U(1)$ gauge field are `dyons' 
(in the Witten sense) with electric charge shifted from integer by $\frac{1}{2}$.  Label particles by $(q_m,q_e)$, where $q_m$ is the magnetic charge (monopole strength) and $q_e$ is the electric charge. A strength-$1$ monopole (dyon) carries charge-$1/2$, labeled as $(1,1/2)$, which under time-reversal transforms to the $(-1,1/2)$ dyon, since electric charge is even while magnetic charge is odd under time-reversal.

Introduce fields $d_{q_m, q_e}$ for dyons with magnetic charge $q_m$ and electric charge $q_e$. Under time reversal these transform as 
\begin{eqnarray}
\mathcal{T}^{-1}d_{(1,1/2)}\mathcal{T}&=&e^{i\alpha}d_{(-1,1/2)} \\ \nonumber
\mathcal{T}^{-1}d_{(-1,1/2)}\mathcal{T}&=&e^{i\beta}d_{(1,1/2)}
\label{dyontrev}
\end{eqnarray}
where $d_{(q_m,q_e)}$ denotes the corresponding dyon operator. 
We then have for $\mathcal{T}^2$
\begin{eqnarray}
\label{t2}
\mathcal{T}^{-2}d_{(1,1/2)}\mathcal{T}^2&=&e^{i(\beta-\alpha)}d_{(1,1/2)} \\ \nonumber
\mathcal{T}^{-2}d_{(-1,1/2)}\mathcal{T}^2&=&e^{i(\alpha-\beta)}d_{(-1,1/2)}
\end{eqnarray}
The exact value of the phase factor $e^{i(\alpha-\beta)}$ is not meaningful since it is not gauge-invariant (see Appendix. \ref{S1}).

Now let's consider the bound state of $d_{(1,1/2)}$ and $d_{(-1,1/2)}$, it has $q_m=0$ and $q_e=1$, which is nothing but the fundamental charge of the system.  
These two dyons see each other as an effective monopole. To see this view the $(-1,1/2)$ dyon as the bound state of the electric charge $(0,1)$ and $(-1,-1/2)$ which is the anti-particle of $(1,1/2)$. The Berry phase seen by the $(-1,1/2)$ dyon is the same as that seen by a charge from a monopole. Hence their bound state will carry half-integer orbital angular momentum and fermionic statistics The angular momentum of the gauge field\cite{witten} in this bound state is given by
\begin{equation}
L=\frac{q_{e,1}q_{m,2}-q_{e,2}q_{m,1}}{2}=1/2.
\end{equation}

The half integer angular momentum goes hand in hand with fermi statistics of the bound state.  To determine whether or not the fermion is a Kramers doublet, we need to consider contributions from the internal and orbital degrees of freedom separately. The internal contribution follows readily from Eq.~\eqref{dyontrev}, which contributes to $\mathcal{T}^2$ by $e^{i(\beta-\alpha)}e^{i(\alpha-\beta)}=1$. The orbital part contributes to $\mathcal {T}^2$ by $-1$ due to the half-integer angular momentum. More precisely, since time-reversal exchanges the two dyons, it is generated by a $\pi$-rotation along a great circle, hence $\mathcal{T}^2$ is generated by a $2\pi$-rotation along a great circle, which picks up a Berry phase of $\pi$ due to the mutual-monopole structure of the two dyons. Therefore we conclude that the fundamental charge must be a Kramers fermion, and there's no fermion SPT with $\theta=\pi$ made out of non-Kramers fermions. We emphasize that this argument is non-perturbative, and does not rely on results from free fermion theories.

In the absence of the $\theta = \pi$ TI for non-Kramers fermions ($T^2 = 1$) what are the possible TIs? The arguments advanced earlier go through as before and we again inherit 
the boson SPTs with symmetry $\mathbb{Z}_2^T$. 
 Thus the classification for interacting non-Kramers fermion TIs with time reversal is $\mathbb{Z}_2^2$. 
 
 Finally we note that the methods of this paper imply the absence of any topological insulator states of electrons when time reversal is absent ({\em i.e} when only charge $U(1)$ is present). Progress toward the classification  of interacting time reversal symmetric electronic topological superconductors  (the charge $U(1)$ is absent)   is made in Ref. \onlinecite{TScSTO} which proposes a $Z_{16}$ classification.  

\begin{figure}[ttt]
\includegraphics[width=2.8in]{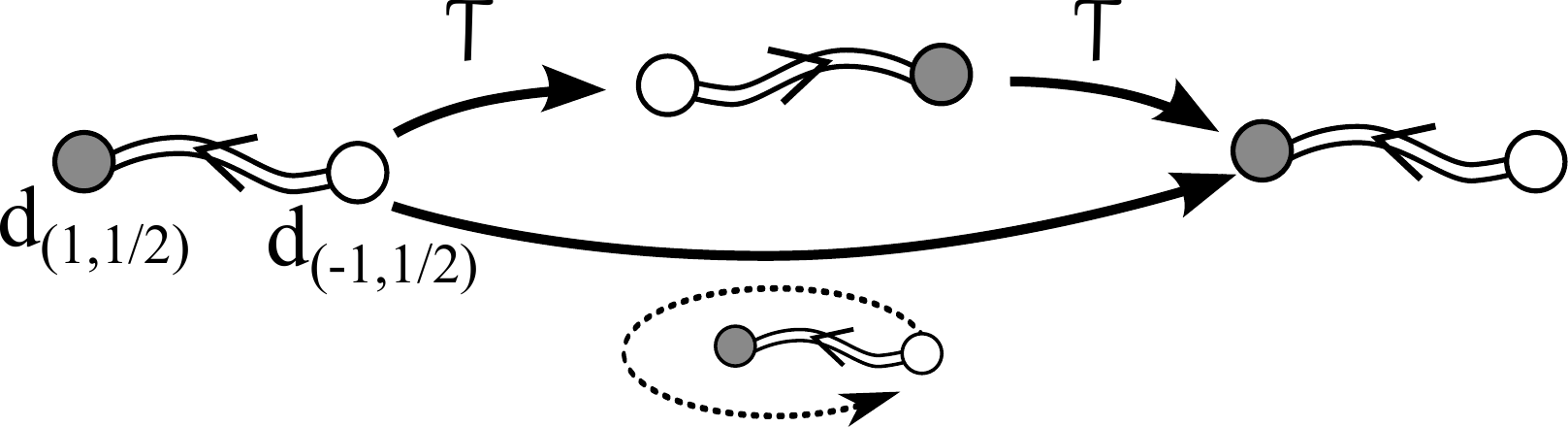}
\caption{For $\theta=\pi$, a monopole and anti-monopole become charge-$\frac{e}{2}$ dyons. Acting twice with $\mathcal{T}$ is equivalent to rotating the pair by $2\pi$, which gives Berry-phase $-1$ due to the half-angular momentum of the EM field of the dyon-pair.}
\label{fig:MonopoleExchange}
\end{figure}

\section{Implications for topological Mott insulators}
\label{S7}
Let us now briefly consider the question of confined phases obtained by condensing the dyons of the topological Mott insulator phase\cite{pesinlb} whose low energy theory is precisely the gauged TI.  Since the $(1,1/2)$ and $(-1,1/2)$ dyons see each other as effective monopoles, they cannot condense simultaneously. Condensing one of them should confine the other, just as condensing monopoles will confine electric charges. Since time-reversal relates these two dyons, this implies that the dyons cannot condense (hence confine the gauge theory) without breaking time-reversal symmetry. That the condensation of either of the $(1, \pm 1/2)$ dyons breaks T-reversal was previously pointed\cite{dcond}. Here we  see that it is not possible to simultaneously condense both dyons. Thus the confined phase obtained from the topological Mott insulator necessarily breaks T-reversal and hence is an antiferromagnet.


\begin{thebibliography}{99}
\bibitem{note1} The `topological Mott insulator'\cite{pesinlb} has exotic gapless bulk excitations.
\bibitem{note2}Such  a phase with coexistence of topological order and superconductivity was denoted SC$^*$ in Ref. \onlinecite{z2long}.
\bibitem{note3}A note for experts: $\nu_Q-\nu$ is determined only up to $8$, so we have $\nu_Q-\nu=0 ({\rm mod} 8)$ for half of the insulators and $\nu_Q-\nu=4({\rm mod} 8)$ for the other half.
\end{thebibliography}
\end{document}